# Ripple formation on Nickel irradiated with radially polarized femtosecond beams

GEORGE D.TSIBIDIS,[1,*] EVANGELOS SKOULAS[1], EMMANUEL STRATAKIS[1,2,*]

[1] *Institute of Electronic Structure and Laser (IESL), Foundation for Research and Technology (FORTH), N. Plastira 100, Vassilika Vouton, 70013, Heraklion, Crete, Greece*
[2] *Materials Science and Technology Department, University of Crete, 71003 Heraklion, Greece*
*Corresponding authors: tsibidis@iesl.forth.gr; stratak@iesl.forth.gr

**We report on the morphological effects induced by the inhomogeneous absorption of radially polarized femtosecond laser irradiation of nickel (Ni) in sub-ablation conditions. A theoretical prediction of the morphology profile is performed and the role of surface plasmon excitation in the production of self-formed periodic ripples structures is evaluated. Results indicate a smaller periodicity of the ripples profile compared to that attained under linearly polarized irradiation conditions. A combined hydrodynamical and thermoelastic model is presented in laser beam conditions that lead to material melting. The simulation results are presented to be in good agreement with the experimental findings. The ability to control the size of the morphological changes via modulating the beam polarization may provide an additional route for controlling and optimizing the outcome of laser micro-processing. © 2015 Optical Society of America**

Laser beams with cylindrical polarization states, namely radial and azimuthal polarization, have gained remarkable attention in the past two decades, as the symmetry of the polarization enables new processing strategies [1] with applications in various fields including microscopy, lithography [2], electron acceleration [3], material processing [1, 4, 5] and optical trapping [6]. The so-called cylindrical vector beams (CVB) have been the topic of numerous theoretical and experimental investigations [4, 7-10].

Surface Plasmon Polariton (SPP) excitation by localized fields originating from CVB and propagation on metal surfaces have been extensively explored [7, 11-14]. While SPP excitation at a metal/air interface is reported to lead to periodic ripple patterns similar to those induced by linearly polarized beams [15-21], only a few studies have been performed to correlate the characteristics of the CVB with induced morphological changes [22, 23]. Nevertheless, a detailed exploration of the physical mechanisms that lead to surface modification induced upon irradiation with CVB (Fig.1) has yet to be considered (i.e. SPP excitation, heat transfer, phase transitions).

In this Letter, we report on the morphological profile (Fig.1a,c), periodicity of laser induced ripples structures (Fig.1b) and damaged area produced upon repetitive irradiation of Ni with femtosecond (fs) pulsed lasers with radial polarization at a laser wavelength of $\lambda_L$=1026nm. Theoretical prediction of the characteristics of the produced structures is performed by solving Maxwell's equations, analysing SPP propagation and computing heat transfer and electron-phonon relaxation processes. The results are compared to the linear polarization case (Fig.1d,e,f). The selection of Ni that is a transition metal with completed $d$ bands was based on its unique thermal properties; specifically, the electron heat capacity of Ni strongly deviates from linear electron temperature dependency, while the electron-phonon coupling $g$ rapidly decreases with increasing electron temperature for low temperatures, followed by a slower decrease as temperature further rises [24]. These properties emphasise the significant role of electron-lattice coupling and electron diffusion in energy confinement and thermalisation that eventually influence size of the laser induced morphological features. We express a radially polarised beam as the superposition of orthogonally Hermite-Gauss $HG_{01}$ and $HG_{10}$ modes [7]

$$\vec{E}_r = HG_{10}\hat{x} + HG_{01}\hat{y} \qquad (1)$$

where $\vec{E}_r$ denotes radial polarization and $\hat{x}$, $\hat{y}$ are the unit vectors along the $x$- and $y$-axis, respectively (Fig.1).

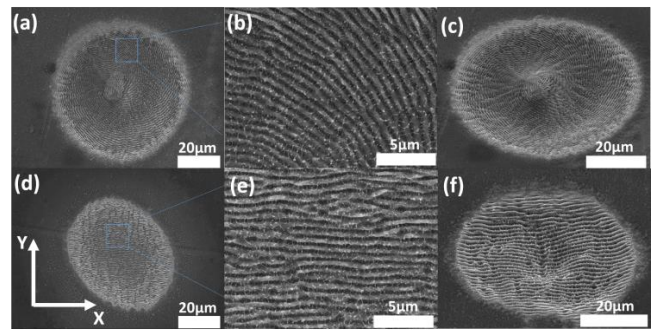

Fig. 1. SEM images of surface profiles after fs-irradiation with radial (a)-(c), and linear polarization of the electric field of the incident beam (d)-(f). $J$=0.36J/cm$^2$, $\tau_p$=170fs, $NP$=200.

Repetitive irradiation of Ni with fs pulses could give rise to excitation of SPP (as the real part of the dielectric constant of Ni is $Re(\varepsilon)$<-1), although the photon energy and the density of states of Ni indicate that interband transitions dominate its optical properties in the infrared range [16]. Corrugated surfaces could also allow SPP excitation [19, 25, 26] and correlation of morphological characteristics of the irradiated zone (corrugation height/amplitude, $\delta$ and grating periodicity, $\Lambda$) with the magnitude of the longitudinal wavevector of the SPP requires a systematic analysis of the propagation of the respective electromagnetic field [19,21]. In this context, the spatio-temporal distribution of the electric field is modelled assuming a surface profile that is determined by the approximating function (where $r=(x^2+y^2)^{1/2}$)

$$\delta(r) = D(r)\sin(2\pi r/\Lambda) \quad (2)$$

In case of a Gaussian laser beam, $D(r) \sim exp(-r^2)$. By contrast, for a CVB, $D(r)$ is determined by the form of the intensity resulting from Eq.1. The interaction between the incident beam and the surface-plasmon wave excited is estimated assuming low modulation of the surface grating structure. Due to the axial symmetry, analysis can be performed on the $r$-$z$ plane, assuming a TM wave for the incident beam. The solution of the Maxwell's equations along with the requirement of the tangential component of the electric field $\vec{E}$ and normal component of $\varepsilon \vec{E}$ on the boundary defined by (2) allows determination of the spatial distribution of the electric field and derivation of the dispersion relations (numerically calculated). To estimate the optimal laser-grating coupling, the combination of maximum height $\delta$ and resonant length $\Lambda$ have been computed that yields enhanced longitudinal $E_r$ inside the irradiation zone (see Appendix). Theoretical results provide a correlation of $\Lambda$ as a function of the ripple depth (Fig.2) that is similar to computations performed for semiconductors [27].

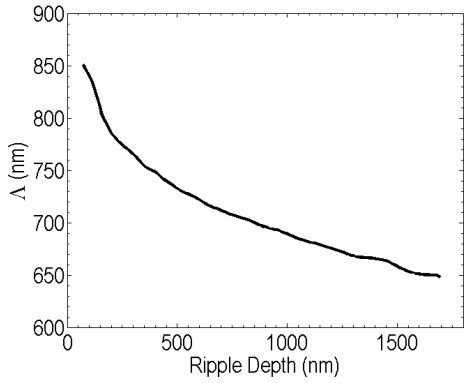

Fig. 2. Dependence of corrugation periodicity on maximum corrugation depth (for RP).

It is evident that $\Lambda$ is a decreasing function of the corrugated surface depth. The electric field spatio-temporal distribution and the associated intensity is incorporated into the source term of a two temperature model (TTM) [28] which is solved to derive the heat transfer, electron-phonon relaxation and heat lattice temperature dynamics. In the model, interband transition contributions are also assumed [29]. Due to the fact that the laser beam conditions (fluence $J$=0.24J/cm$^2$, pulse duration $\tau_p$=170fs, spot diameter=34μm) are sufficient to melt partially the material, a Navier-Stokes equation to predict the movement of the produced molten material is used [19]; furthermore, the movement of a portion of the material which does not undergo melting but is subjected to stress fields is described by a thermoelastic model [30]. This combined hydrodynamic/elasto-plastic model is used to describe the induced surface profile. In particular, the calculated lattice temperature field evolution determines the flow dynamics of the molten material and predicts plastic effects. The following set of differential equations will be used to describe: (i) heat absorption, heat transfer and relaxation processes (first two), (ii) fluid dynamics (third and fourth), (iii) elastoplastic effects (fifth, sixth and seventh)

$$C_e \frac{\partial T_e}{\partial t} = \vec{\nabla} \cdot (\kappa_e \vec{\nabla} T_e) - g(T_e - T_L) + S(\vec{r},t)$$

$$C_L \frac{\partial T_L}{\partial t} = g(T_e - T_L) - (3\lambda_{Ni} + 2\mu)\alpha' T_L \sum_{i=1}^{3} \dot{\varepsilon}_{ii}$$

$$C_L \left( \frac{\partial T_L}{\partial t} + \vec{\nabla} \cdot (\vec{u} T_L) \right) - L_m \delta(T_L - T_{melt}) \frac{\partial T_L}{\partial t} = \vec{\nabla} \cdot (\kappa_L \vec{\nabla} T_L)$$

$$\rho_L^{(m)} \left( \frac{\partial \vec{u}}{\partial t} + \vec{u} \cdot \vec{\nabla} \vec{u} \right) = \vec{\nabla} \cdot \left( -P + \mu^{(m)}(\vec{\nabla} \vec{u}) + \mu^{(m)}(\vec{\nabla} \vec{u})^T \right) \quad (3)$$

$$\rho_L^{(s)} \frac{\partial^2 V_i}{\partial t^2} = \sum_{j=1}^{3} \frac{\partial \sigma_{ji}}{\partial x^j}, \quad i,j=1,2,3$$

$$\sigma_{ij} = 2\mu\varepsilon_{ij} + \lambda_{Ni}\sum_{k=1}^{3}\varepsilon_{kk}\delta_{ij} - \delta_{ij}(3\lambda_{Ni}+2\mu)\alpha'(T_L - 300)$$

$$\varepsilon_{ij} = 1/2\left( \frac{\partial V_i}{\partial x^j} + \frac{\partial V_j}{\partial x^i} \right)$$

where $T_e$ and $T_L$ are the electron and lattice temperatures, $k_e$ and $k_L$ stand for the electron and lattice heat conductivities, $C_e$ and $C_L$ correspond to the electron and lattice heat capacities, $\mu^{(m)}$ is the dynamic viscosity of the liquid, $\rho^{(m)}$ and $\rho^{(s)}$ are the material densities in molten and solid phase, respectively. Finally, $\alpha'$ is the thermal expansion coefficient and $P$ is the pressure. On the other hand, $\sigma_{ij}$ and $\varepsilon_{ij}$ stand for stress and strains, while $\lambda_{Ni}$ and $\mu$ are the Lame's coefficients. $V_i$ are the displacements along the $x_i$ direction while $\vec{u}$ describes the fluid velocity. $L_m$ is the latent heat of fusion of Ni and $S(\vec{r},t)$ is the source term due to laser. The thermophysical or thermoelastic parameters that appear in the equations are derived using fitting techniques [22] or bibliographical records [29-31]. Fig.3a shows the spatial distribution of the lattice temperature at $t$=20ps after the material has been irradiated with two identical, 170fs, pulses at a fluence of 0.24J/cm$^2$.

Simulation results demosntrate that no ripples are formed for single ($NP$=1) pulse irradiation as SPP excitation is not possible [19]. The corresponding surface displacement at time $t$=0.45ns is illustrated in Fig.3b. The resolidification analysis, determined by following the evolution of the isothermal $T_{melt}$, allows calculation of the corrugation amplitude. The theoretical calculations indicate that the produced surface profile upon resolidification is characterized not only by the formation of subwavelength ripples but also by a cone protruding along the $z$-axis (Fig. 3b). The latter is found to be formed by stress fields, rather than fluid movement, since the induced lattice temperature in the vicinity of the spot centre is not sufficient to melt the material. In particular, it is observed that the radius of the produced 'Mexican-hat' is determined by the magnitude of horizontal strain components while the peak of the cone is formed below the level of the initially flat surface. The theoretical predictions (Fig.3b) on such produced morphology comply with the experimental observations (inset in Fig.3c). We have to emphasise that although there is an evident ablation in the presented SEM image taken for $NP$=2000 (Fig.3c), a similar morphological profile is obtained for irradiation with fewer number of pulses ($NP$=2) where only a minimal mass removal occurs (subablation conditions [19]).

In order to derive a comprehensive quantitative analysis of the morphological effects induced by radially polarized fs beams, we explore the dependence of the surface modification on the laser fluence and the number of incident pulses, assuming also the unique behaviour of $g$. While the fluence increase leads to a rise in the electron temperature $T_e$ (which is periodically modulated as a result of the initial SPP-originated periodical non-uniform energy deposition), the decrease of $g$ with increasing $T_e$ leads to a delay of the heat transfer to the lattice via electron-phonon coupling with an increased efficiency of energy localization [31]. Therefore, the initial $T_e$ periodic distribution is expected to weaken substantially prior to heat transfer to the lattice, which will result to a less pronounced heat distribution to the lattice and eventually to a smaller ripple height (see for example Fig.4, for *NP*=100).

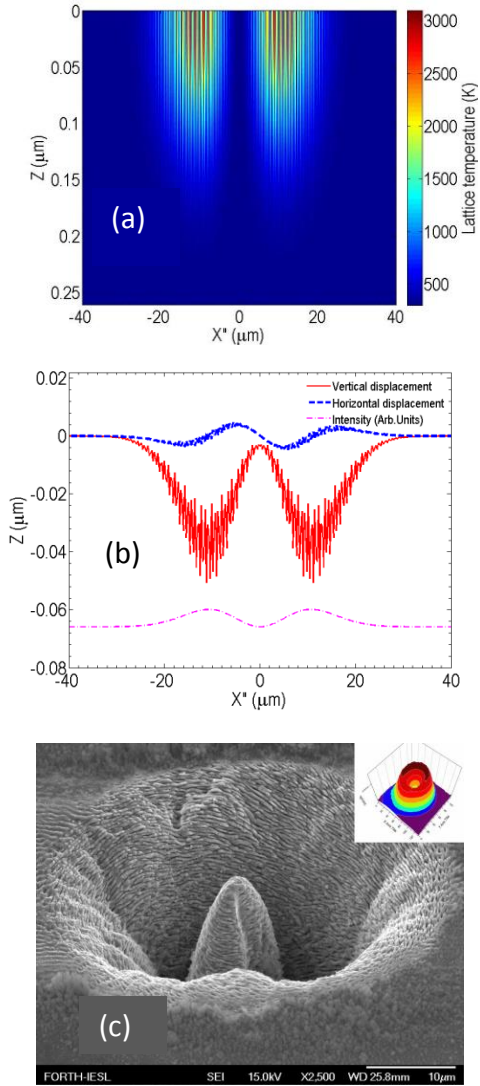

Fig. 3. (a) Temperature profile at *t*=20ps for a RP beam ($X'' \in [-r, r]$ and it is the axis along the direction of the unit vector $\hat{r} = (\hat{x} + \hat{y})/\sqrt{2}$ ), (b) Vertical and horizontal displacements at *t*=0.45ns for *NP*=2, (c) SEM image of surface profile with radially polarized beam at 2000 pulses; inset indicates the profile (*J*=0.24J/cm$^2$, $\tau_p$=170fs).

Fig.4 also illustrates a comparison of ripple height variation produced by radially (RP) and linearly (LP) polarized beams at various fluences respectively. It is evident that due to larger energy deposition for radially polarized light, the maximum ripple depth is always larger than the one expected from a linearly polarized light. Furthermore, the larger heat diffusion to higher depths due to increased lattice temperatures for RP, gives rise to propagation of the damage in larger depths. Calculations were performed in a range of fluences between the melting and ablation thresholds (~120mJ/cm$^2$ and ~400mJ/cm$^2$, respectively).

To test the above theoretical predictions, an appropriate experimental protocol is developed. Polished Ni films of 99.9% purity and average thickness of 150μm were used. An Yb:KGW laser was used to produce 170fs pulses, 1 KHz repetition rate and 1026nm central wavelength. Beam polarization was transformed from linear to radial using an optical vortex polarization converter [10]. The irradiation pulses were focused onto the sample by a spherical N-BK7 lens (60mm focal length), while the FWHM of the spot diameter was 34μm. Two dimensional Fast Fourier Transform algorithms were used on SEM images to compute the ripple periodicity.

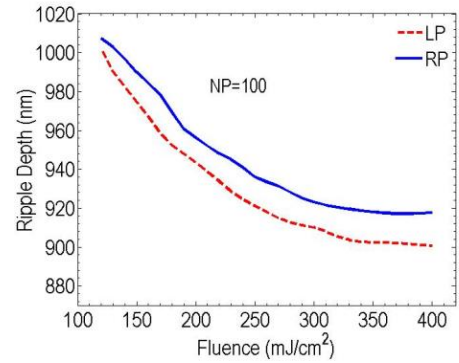

Fig. 4. Maximum ripple depth *vs.* fluence for RP/LP.

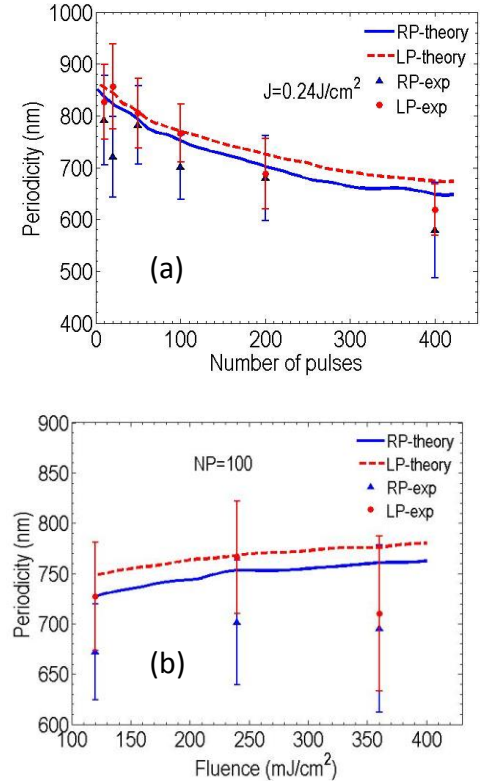

Fig. 5. (Color Online). Comparison of ripple periodicity *vs. NP* (a) and (b) fluence for RP and LP beams. Both the experimental observations and theoretical predictions are presented in each case.

A comparison of the theoretical predictions with the experimental results is illustrated in Fig.5 for both the RP and the LP case, under identical irradiation conditions. It is observed that irradiation with a RP beam results in a reduced ripple periodicity compared to that derived using a LP beam. The respective simulations indicate that this deviation can be attributed to both electrodynamical and hydrodynamical effects. Specifically, as the local energy deposition on the material is higher and is diffused to larger depths for RP, it produces an increased maximum of the $E_r$ at smaller $\Lambda/\lambda$ (for subsequent irradiation), which means that the grating wavelength will decrease. Besides this, RP results in higher lattice temperature which, in turn, leads to enhanced fluid vortices development that further decrease the ripple periodicity [19]. Results on the pulse number dependence of ripple periodicity demonstrate a good agreement between the tendency of the experimental observations and theoretical predictions (Fig.5a). By contrast, for both types of polarization, the irradiation fluence does not appear to influence substantially the ripple periodicity (Fig..5b).

In conclusion, we have performed a systematic, theoretical approach complemented with experimental studies to interpret the surface profile and the periodicity of the self-assembled ripples formed upon irradiation of Ni with RP fs laser pulses. The study demonstrates the significant influence of the incident beam polarization on both the morphological profile as well as the size of the produced structures. Despite the exploration of morphological effects on metals, the study can also be extended to other types of materials, including dielectrics, semiconductors or polymers. The ability to control the size of the morphological changes via modulating the beam polarization may provide novel types of surface and bulk structures with significant advantages for potential applications.

**Funding.** *LiNaBiofluid* (funded by EU's H2020 framework programme for research and innovation under Grant Agreement No. 665337) and N*anoscience Foundries and Fine Analysis* (NFFA)–Europe H2020-INFRAIA-2014-2015 (Grant agreement No 654360) and *SOLAR-NANO* (Greek-German)*;* projects for financial support.

## Appendix A: Computation of Electric field

To provide a quantitative description of the role of the SP excitation in determining the surface profile of the irradiated material, a systematic analysis is pursued based on the propagation of the electric field that is developed on the surface of the material. To this end, we solve the Maxwell equations, assuming a TM wave (magnetic field component is perpendicular to the *r-z* plane). We consider a 3D Cartesian coordinate system defined by *X"-Y"-z* , where *Y"* is perpendicular to the *X"-z* (or *r-z*) plane. The electric and magnetic fields are

$$\vec{E}^{(j)} = \begin{pmatrix} E_{X"}^{(j)} \\ 0 \\ E_{z}^{(j)} \end{pmatrix} e^{-i\omega t} e^{ik_{X"}^{(j)} X" - k_{z}^{(j)} z}$$

$$\vec{H}^{(j)} = \begin{pmatrix} 0 \\ H_{Y"}^{(j)} \\ 0 \end{pmatrix} e^{-i\omega t} e^{ik_{X"}^{(j)} X" - k_{z}^{(j)} z}$$

(A.1)

where the subscript *j* (is + or -) to indicate that it refers to regions above (A) or below (B), respectively, the separating line that defines the surface morphology (see Eq.2). In Eq.A.1, $\omega$ is the angular frequency to be determined from the dispersion relations, $k_{X"}^{(+)} = k_{X"}^{(-)} \equiv k_{X"}$ is the component of the wavevector of the surface wave along the *X"*-axis and

$$k_{z}^{(j)} = \left( \left(k_{X"}^{(j)}\right)^2 - \frac{\omega^2}{c^2} \varepsilon^{(j)}(\omega) \right)^{1/2}$$

(A.2)

In Eq.A.2, $\varepsilon^{(j)}$ is the dielectric constant in regions A (it is assumed that the medium is air which implies $\varepsilon^{(A)}$ =1) and B (assume also contributions form interband transitions [29]), respectively. At the interface of the produced grating, the boundary conditions that require the tangential component of the electric field and the normal component of $\vec{D} = \varepsilon(\omega)\vec{E}$ are continuous. Thus, Eqs.2, A.1, A.2 and

$$E_{//}(X'',z) = \left( E_{X''}(X'',z) + \frac{d\delta(X'')}{dX''} E_z(X'',z) \right) \left[ \frac{(dX'')^2}{(dX'')^2 + (d\delta(X''))^2} \right]^{1/2}$$

$$D_{\perp}(X'',z) = \left( D_{X''}(X'',z) - \frac{d\delta(X'')}{dX''} D_{X''}(X'',z) \right) \left[ \frac{(dX'')^2}{(dX'')^2 + (d\delta(X''))^2} \right]^{1/2}$$

(A.3)

enable determination of the spatial distribution of the electric field everywhere near the surface of the material.

**Supporting Material**

**Appendix A: Spatial distribution of the laser beam**

The spatial distribution of the two beams (RP and LP) along a cross line on the *x-y* plane that passes through the origin are shown in Fig.A1. The two profiles are assumed to correspond to the same deposited total energy.

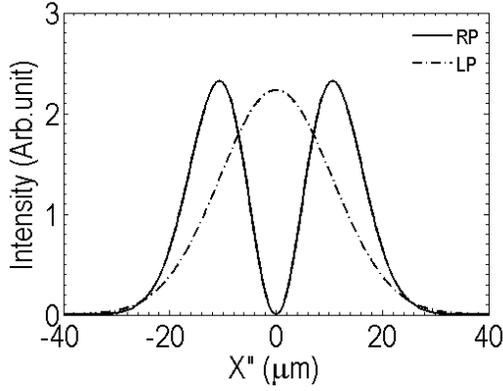

Fig. A1. Spatial intensity distribution of Radially (RP) and Linearly (LP) polarized beam. $X'' \in [-r,r]$ and it is the axis along the direction of the unit vector $\hat{r}$, (i.e. $\hat{r} = (\hat{x} + \hat{y})/\sqrt{2}$).

**Appendix B: Simulation parameters**

| Thermophysical and thermomechanical properties | | | |
|---|---|---|---|
| Parameter | Units | Value | Reference |
| $C_e$ | $10^5$ J/m$^3$K | Fitting | [24,33] |
| $A$ | $10^7$s$^{-1}$K$^{-2}$ | 0.59 | [32] |
| $B$ | $10^{11}$s$^{-1}$K$^{-1}$ | 1.4 | [32] |
| $k_e$ | Jm$^{-1}$s$^{-1}$K$^{-1}$ | $318*B*T_e/(A*T_e^2+BT_L)$ | [32] |
| $g$ | $10^{17}$ Wm$^{-3}$K$^{-1}$ | Fitting | [24,33] |
| $c_L^{(m)}$ | J Kgr$^{-1}$ K$^{-1}$ | 734.16 | [31] |
| $c_L^{(s)}$ | J Kgr$^{-1}$ K$^{-1}$ | For 298 K<$T_L$<1400: $-295.95+4.95T_L-0.0096(T_L)^2+$ $9.46\times10^{-6}(T_L)^3$ $-4.36\times10^{-9}(T_L)^4+7.6\times10^{-13}(T_L)^5$ For $T_L$>1400: 616.56 | [31] |
| $Y$ | GPa | 200 | [31] |
| $P$ | - | 0.31 | [31] |
| $\lambda_{2\delta}=\frac{Y*P}{(1+P)*(1-2*P)}$ | GPa | $1.24*10^8$ | |
| $\mu=\frac{Y}{2*(1+P)}$ | GPa | $7.63*10^7$ | |
| $\alpha'$ | $10^{-6}$K$^{-1}$ | 13.4 | [31] |
| $L_m$ | kJ/Kgr | 297 | [31] |
| $k_L$ | W(mK)$^{-1}$ | For 298 K<$T_L$<600: $114.43-0.082T_L$ For $T_{melt}$>$T_L$>600: $50.47+0.012T_L$ For $T_L$>$T_{melt}$: 89 | [31] |
| $T_{melt}$ | K | 1728 | [31] |
| $\rho_L^{(m)}$ | gr/cm$^3$ | 7.75 | [31] |
| $\rho_L^{(s)}$ | gr/cm$^3$ | 8.9 | [31] |

Visualisation 3. Simulation parameters

Fig. A2. Simulation parameters

**Appendix C: Energy heat transfer and phase change**

To compute the energy absorption, heat transfer and relaxation process the temperature dependence of the thermophysical parameters is considered [32]. A minimal mass removal (i.e. we call it 'subablation conditions' to distinguish effects resulting from the normal 'ablation conditions' that assume plume formation, presence of ejected fragments, etc) [19]. Phase change and resolidification process is incorporated into the model by solving the Navier-Stokes equation. A finite-difference method in a staggered grid is employed [19] to solve numerically the heat transfer equations (i.e. by solving the differential equations that describe the two temperature model) [19], phase change and dynamic elasticity equations [28]. Plastic effects are considered when the total stress exceeds the yield stress of the material [28].

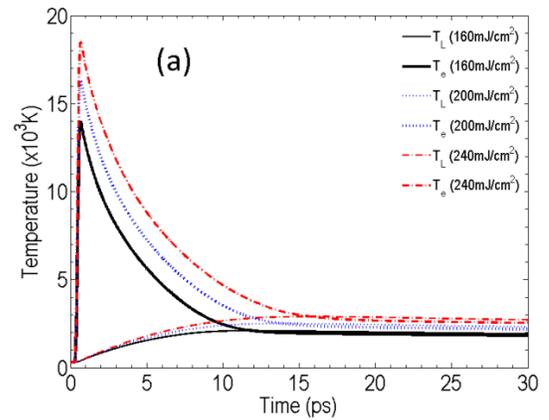

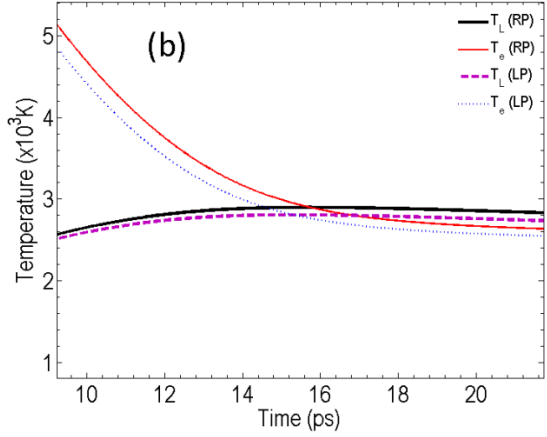

Fig. A3. (Color Online). (a) Evolution of maximum lattice and electron temperatures for various fluences, for RP beam. (b) comparison of electron and lattice temperatures for RP and LP beams for $J$=0.24J/cm² ($NP$=1, $\tau_p$=170fs).

Simulation results for the electron and lattice temperatures show the evolution of temperatures for various fluences (for melting conditions). It is evident that (i) larger fluences lead to a delayed relaxation time (due to the electron-phonon coupling temperature dependence) as seen in Fig.A3.a, (ii) Radial polarization produces slightly higher temperatures and delay of the electron-phonon relaxation (Fig.A3.b).

**Appendix D: Morphological changes/Displacement**

  a. *$NP$=1*

Fig.A4.a illustrates the temperature spatial lattice temperature variation at t=20ps that results from consideration of SPP excitation for $NP$=1 for $J$=0.24J/cm². Fig.A4.b shows at t=0.45ns the horizontal and vertical displacements. To emphasise on the fluid movement, focus of the forces that induce material movement has been restricted. For $NP$=1 (for which, it is important to note that momentum conservation violation does not allow SPP excitation) as surface-tension generated forces influence similarly material movement in every periodically situated ripple well.

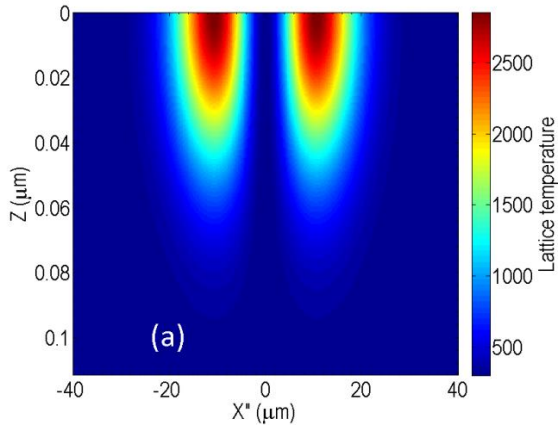

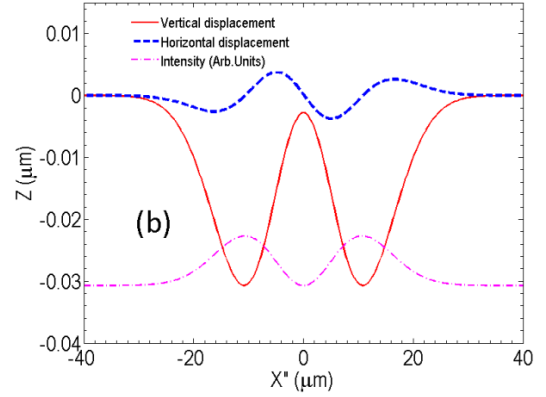

Fig. A4. (Color Online). (a) Temperature profile at $t$=20ps for a RP beam, (b) Vertical and horizontal displacements at $t$=0.45ns. ($J$=0.24J/cm², $NP$=1, $\tau_p$=170fs).

  b. *$NP$=2*

A similar analysis is performed for $NP$=2 (when rippled structures are formed as a result of SPP excitation). Fig.A5 illustrates the temperature profile and the vertical and horizontal displacements.

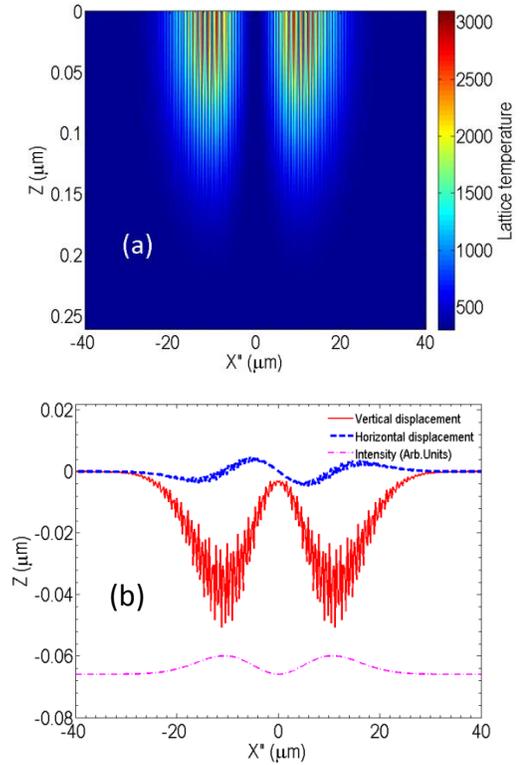

Fig. A5. (Color Online). (a) Temperature profile at $t$=20ps for a RP beam, (b) Vertical and horizontal displacements at $t$=0.45ns. ($J$=0.24J/cm², $NP$=2, $\tau_p$=170fs).

**Appendix E: Experimental setup**

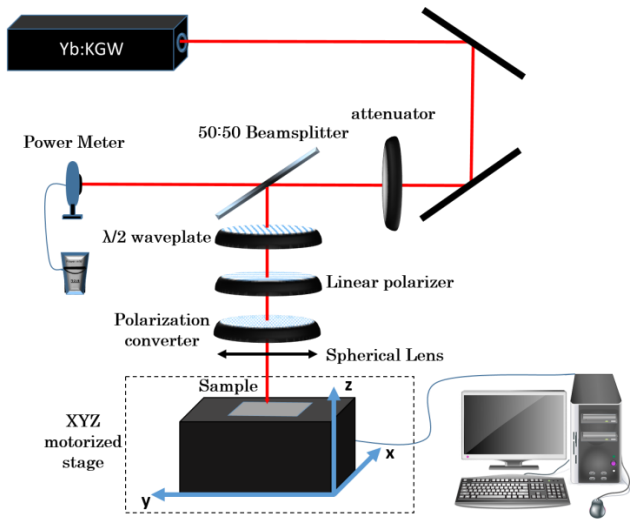

Fig. A6. Experimental setup.

Polished Ni films of 99.9% purity and average thickness 150μm were used (Fig.A6). An Yb:KGW laser was used to produce linearly polarized pulses of pulse duration equal to 170fs, 1 KHz repetition rate and 1026nm central wavelength. Beam polarization was transformed from linear to radial with an optical vortex polarization converter structured glass [10]. Pulses were focused with a spherical N-BK7 lens (60mm focal length) and spot diameter was 26μm (at FWHM) measured by a CCD camera. Samples were positioned perpendicularly to the incident beam and all experiments were performed in air on a 3-axis motorized stage. Two different types of experiments were conducted to correlate characteristics of the surface modification (i.e. periodicity of ripples, damage size, spot area) with the laser beam polarization (linear or radial), number of pulses and fluence. Surface images were analysed by SEM and AFM and ripple periodicity was computed by using a 2D fast Fourier transform on SEM images. As explained in the main text, while the damage induced on the material is comparable for the two polarisations, the horizontal movement of the affected zone produces remarkable morphological changes leading to the formation of a 'Mexican hat'-shaped area. For RP, the enhanced horizontal movement due to larger temperature gradients along that direction causes a squeezing effect to the 'cone' that is produced and eventually the radius of the cone decreases with increasing fluence (Fig.A7).

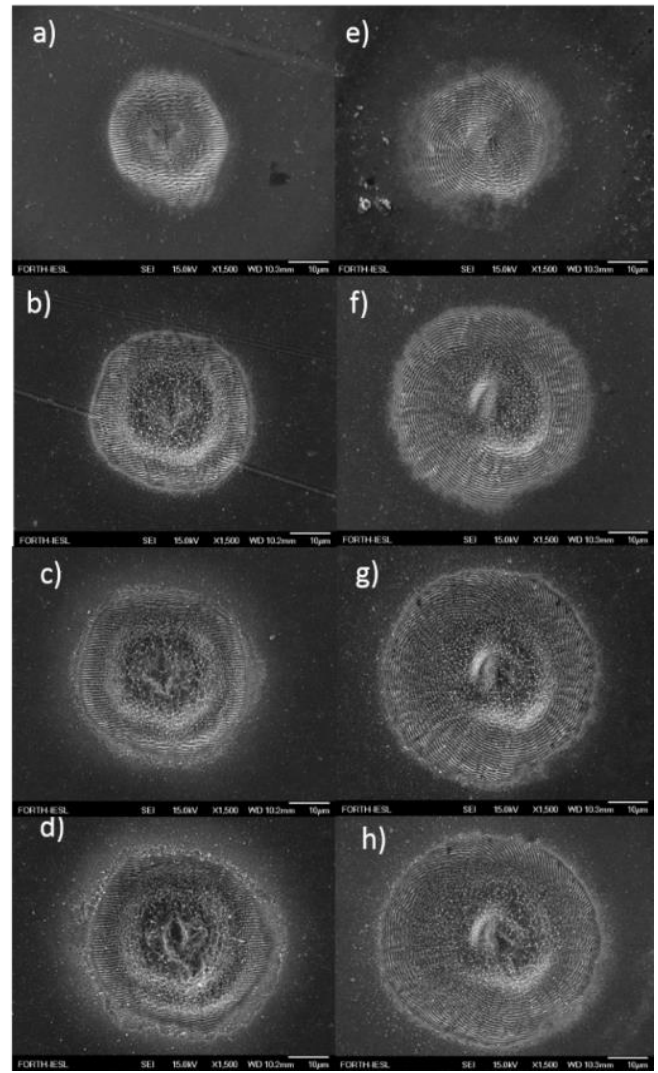

Fig. A7. SEM images of 100 pulses on Ni with LP (a-d) and RP (e-h) beam with fluences of 0.12–0.36-0.59-0.84 J/cm$^2$.

## Appendix F: Experimental results

The two types of laser beam polarisation do not produce remarkable alterations in the depth of the corrugated profile for increasing number of pulses. Nevertheless, feedback and hydrodynamical effects lead to larger ripple well depth that is associated also with a decreasing $\Lambda$ as explained in the main text (Fig.A8).

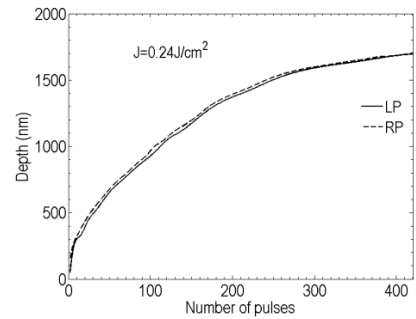

Fig. A8. Maximum ripple well depth *vs.* number of pulses and fluence for RP/LP.